\documentclass[showpacs]{revtex4}
\usepackage{indentfirst}
\begin{document}
\title{\bf Relativistic quark-gluon description of $ ^3 He$}
\author{S.M. Gerasyuta}
\email{gerasyuta@SG6488.spb.edu}
\author{E.E. Matskevich}
\email{matskev@pobox.spbu.ru}
\affiliation{Department of Theoretical Physics, St. Petersburg State University, 198904,
St. Petersburg, Russia}
\affiliation{Department of Physics, LTA, 194021, St. Petersburg, Russia}
\begin{abstract}
The relativistic nine-quark equations are found in the framework of the
dispersion relation technique. $ ^3 He$ nucleus is described by these
equations. We consider the $ ^3 He$ as the system of interacting quarks
and gluons. The approximate solutions of these equations using the method
based on the extraction of leading singularities of the amplitudes are
obtained. The relativistic nine-quark amplitudes of $ ^3 He$,
including the $u$, $d$ quarks are calculated. The poles of these amplitudes
determine the mass of nine-quark system. The $ ^3 He$ mass $M=2809\, MeV$
is calculated. The gluon coupling constant in the light nuclei region
is equal to $g=0.1536$. The gluon interaction of $ ^3 He$ is obtained
in 2 -- 3 time smaller as compared with baryon interaction.
\end{abstract}
\maketitle
\section{Introduction.}
In nuclear physics one studies interacting hadrons in the nonperturbative
regime of quantum chromodynamics (QCD). This field faces two natural
frontiers. The first one is devoted to the origin and the fundaments
of the strong force between the nucleous, as well as its parametrization
in terms of effective degrees of freedom, suitable for the nonperturbative
regime. The other is the completely microscopic, quantum mechanical
treatment of many-body systems (nuclei) in terms of these effective
degrees of freedom with the strong force as dynamical input. Within these
two frontiers, few-nucleon physics at low energy represents quite a rich
and flourishing research field describing a greate variety of strong
interaction phenomena.

The study of polarization phenomena in hadron and hadron-nucleus interactions
gives more detailed information on dynamics of their interactions and the
structure of colliding particles. The quark structure and relativistic
effects of light nuclei, in particular, deuterons, is one of the important
problems in nuclear physics at low and intermediate energies. The theoretical
and experimental study of reactions like the elastic $e-d$ \cite{1} and
$p-d$ \cite{2, 3} scattering, deuteron break-up reactions induced by electrons
or protons \cite{4, 5} and the deuteron stripping processes on protons and
nuclei at the low and intermediate energies \cite{6, 7}, can allow us to
find out new information on the deuteron structure at short distances.
The elastic backward proton-deuteron scattering has been experimentally
and theoretically studied in Saclay \cite{2}, Dubna and at JLab \cite{8, 9, 10, 11}.
Usually these processes are analyzed within a simple impulse approximation.
Up to now all these data have not been described within the one-nucleon
exchange model (ONE), including even the relativistic effects in the
deuteron \cite{12, 13}. Then the elastic backward proton-deuteron scattering
within the relativistic approach including the ONE and the high order graphs
corresponds to the emission, rescattering and absorption of the virtual pion
by a nucleon of deuteron \cite{14}.

The main aim of the investigation of the reaction by the polarized deuterons
is to establish the nature of $2N$ and $3N$ forces, the role of the relativistic
effects and nonnucleon degrees of freedom. The last decades such investigation
were performed at different experiments all over the world at RIKEN, KVI, IUCF
and RCNP. The activing was stimulated by the discrepancy of 30\% between the
measured cross section for deuteron-proton elastic scattering at intermediate
energies and the Faddeev calculations using modern potentials of nucleon-nucleon
interaction.

In Ref. \cite{15} a representation of the Faddeev equation in the form of a
dispersion relation in the pair energy of the two interacting particles
was used. This was found to be convenient in order to obtain an approximate
solution of the Faddeev equation by a method based on extraction of leading
singularities of the amplitude. With a rather crude approximation of the
low-energy $NN$ interaction a relatively good description of the form factor
of tritium (helium-3) at low $q^2$ was obtained.

In our papers \cite{16, 17, 18} relativistic generalization of the three-body
Faddeev equations was obtained in the form of dispersion relations in the
pair energy of two interacting quarks. The mass spectrum of $S$-wave
baryons including $u$, $d$, $s$ quarks was calculated by a method based on
isolating the leading singularities in the amplitude. We searched for the
approximate solution of integral three-quark equations by taking into
account two-particle and triangle singularities, all the weaker ones being
neglected. If we considered such an approximation, which corresponds to
taking into account two-body and triangle singularities, and defined all
the smooth functions of the subenergy variables (as compared with the
singular part of the amplitude) in the middle point of the physical region
of Dalitz-plot, then the problem was reduced to the one of solving a system
of simple algebraic equations.

In the recent paper \cite{19} the relativistic six-quark equations are found
in the framework of coupled-channel formalism. The dynamical mixing between
the subamplitudes of hexaquark are considered. The six-quark amplitudes
of dibaryons are calculated. The poles of these amplitudes determine the
masses of dibaryons. We calculated the contribution of six-quark
subamplitudes to the hexaquark amplitudes.

In the present paper the $ ^3 He$ as the system of interacting quarks and
gluons is considered. The relativistic nine-quark equations are found
in the framework of the dispersion relation technique. The dynamical
mixing between the subamplitudes of $ ^3 He$ is taken into account.
The relativistic nine-quark amplitudes of $ ^3 He$, including the
$u$, $d$ quarks are calculated. The approximate solutions
of these equations using the method based on the extraction of leading
singularities of the amplitude are calculated. The poles of the nine-quark
amplitudes determine the mass of $ ^3 He$. The mass of $ ^3 He$ state is
equal to $M=2809\, MeV$. The model use only two parameters:
cutoff $\Lambda=11$ \cite{19} and gluon coupling constant $g=0.1536$,
which is smaller then the gluon coupling constant of hadron physics.

In Sec. II, we briefly discuss
the relativistic Faddeev approach. The relativistic three-quark equations
are constructed in the form of the dispersion relation over the two-body
subenergy. The approximate solution of these equations using the method
based on the extraction of leading singularities of the amplitude are
obtained. We calculated the mass spectrum of $S$-wave baryons with
$J^P=\frac{1}{2}^+$, $\frac{3}{2}^+$ (Table \ref{tab1}).

In Sec. III, the nine-quark amplitudes of $ ^3 He$ are constructed.
The dynamical mixing between the subamplitudes of $ ^3 He$ is considered.
The relativistic nine-quark equations are constructed in the form of the
dispersion relation over the two-body subenergy. The approximate solutions
of these equations using the method based on the extraction of leading
singularities of the amplitude are obtained.

Sec. IV is devoted to the calculation results for the $ ^3 He$ ($ ^3 H$).

In conclusion, the status of the considered model is discussed.

\section{Brief introduction of relativistic Faddeev equations.}

We consider the derivation of the relativistic generalization of the
Faddeev equation for the example of the $\Delta$-isobar
($J^P=\frac{3}{2}^+$). This is convenient because the spin-flavour part
of the wave function of the $\Delta$-isobar contains only nonstrange quarks
and pair interactions with the quantum numbers of a $J^P=1^+$ diquark
(in the color state $\bar 3_c$). The $3q$ baryon state $\Delta$ is
constructed as color singlet. Taking into account the equality of all pair
interactions of nonstrange quarks in the state with $J^P=1^+$, we obtain the
corresponding equation for the amplitudes (Fig. 1):

\begin{equation}
\label{1}
A_1 (s, s_{12}, s_{13}, s_{23})=\lambda+A_1 (s, s_{12})+
A_1 (s, s_{13})+A_1 (s, s_{23})\, . \end{equation}

\noindent
Here, the $s_{ik}$ are the pair energies of particles 1, 2 and 3, and $s$
is the total energy of the system. To write down a concrete equations for the
function $A_1 (s, s_{12})$ we must specify the amplitude of the pair interaction
of the quarks. We write the amplitude of the interaction of two quarks in
the state $J^P=1^+$ in the form:

\begin{equation}
\label{2}
a_1(s_{12})=\frac{G^2_1(s_{12})}
{1-B_1(s_{12})} \, ,\end{equation}

\begin{equation}
\label{3}
B_1(s_{12})=\int\limits_{4m^2}^{\infty}
\, \frac{ds'_{12}}{\pi}\frac{\rho_1(s'_{12})G^2_1(s'_{12})}
{s'_{12}-s_{12}} \, ,\end{equation}

\begin{eqnarray}
\label{4}
\rho_1 (s_{12})&=&
\left(\frac{1}{3}\, \frac{s_{12}}{4m^2}+\frac{1}{6}\right)
\left(\frac{s_{12}-4m^2}{s_{12}}\right)^{\frac{1}{2}} \, .
\end{eqnarray}

\noindent
Here $G_1(s_{12})$ is the vertex function of a diquark with $J^P=1^+$.
$B_1(s_{12})$ is the Chew-Mandelstam function \cite{20}, and $\rho_1 (s_{12})$
is the phase spaces for a diquark with $J^P=1^+$.

The pair quarks amplitudes $qq \to qq$ are calculated in the framework of
the dispersion $N/D$ method with the input four-fermion interaction
\cite{21, 22, 23} with the quantum numbers of the gluon \cite{24, 25}.

The four-quark interaction is considered as an input:

\begin{eqnarray}
\label{5}
 & g_V \left(\bar q \lambda I_f \gamma_{\mu} q \right)^2 +
2\, g^{(s)}_V \left(\bar q \lambda I_f \gamma_{\mu} q \right)
\left(\bar s \lambda \gamma_{\mu} s \right)+
g^{(ss)}_V \left(\bar s \lambda \gamma_{\mu} s \right)^2
 \, . & \end{eqnarray}

\noindent
Here $I_f$ is the unity matrix in the flavor space $(u, d)$. $\lambda$ are
the color Gell-Mann matrices. Dimensional constants of the four-fermion
interaction $g_V$, $g^{(s)}_V$ and $g^{(ss)}_V$ are parameters of the
model. At $g_V =g^{(s)}_V =g^{(ss)}_V$ the flavor $SU(3)_f$ symmetry occurs.
The strange quark violates the flavor $SU(3)_f$ symmetry. In order to
avoid additional violation parameters we introduce the scale of the
dimensional parameters \cite{25}:

\begin{eqnarray}
\label{6}
g=\frac{m^2}{\pi^2}g_V =\frac{(m+m_s)^2}{4\pi^2}g_V^{(s)} =
\frac{m_s^2}{\pi^2}g_V^{(ss)} \, ,
\quad\quad\quad
\Lambda=\frac{4\Lambda(ik)}
{(m_i+m_k)^2} \, . \end{eqnarray}

\noindent
Here $m_i$ and $m_k$ are the quark masses in the intermediate state of
the quark loop. Dimensionless parameters $g$ and $\Lambda$ are supposed
to be constants which are independent of the quark interaction type.

In the case under consideration the interacting pairs of particles do not
form bound states. Thefore, the integration in the dispersion integral (\ref{7})
run from $4m^2$ to $\infty$. The equation corresponding to Fig. 1 can be
written in the form:

\begin{eqnarray}
\label{7}
A_1(s,s_{12})&
=&\frac{\lambda_1 B_1(s_{12})}{1-B_1(s_{12})}+
\frac{G_1(s_{12})}{1-B_1(s_{12})}
\int\limits_{4m^2}^{\infty}
\, \frac{ds'_{12}}{\pi}\frac{\rho_1(s'_{12})}
{s'_{12}-s_{12}}G_1(s'_{12})\nonumber\\
&&\nonumber\\
 & \times & \int\limits_{-1}^{+1} \, \frac{dz}{2}
[A_1(s,s'_{13})+A_1(s,s'_{23})] \, .
\end{eqnarray}

In Eq. (\ref{7}) $z$ is the cosine of the angle between the relative momentum
of particles 1 and 2 in the intermediate state and the momentum of the third
particle in the final state in the c.m.s. of the particles 1 and 2. In
our case of equal mass of the quarks 1, 2 and 3, $s'_{13}$ and $s'_{12}$
are related by the equation \cite{26}

\begin{eqnarray}
\label{8}
s'_{13}=2m^2-\frac{(s'_{12}+m^2-s)}{2}
\pm \frac{z}{2} \sqrt{\frac{(s'_{12}-4m^2)}{s'_{12}}
(s'_{12}-(\sqrt{s}+m)^2)(s'_{12}-(\sqrt{s}-m)^2)}\, .
\end{eqnarray}

The expression for $s'_{23}$ is similar to (\ref{8}) with the replacement
$z\to -z$. This makes it possible to replace
$[A_1(s,s'_{13})+A_1(s,s'_{23})]$ in (\ref{7}) by $2A_1(s,s'_{13})$.

From the amplitude $A_1(s,s_{12})$ we shall extract the singularities
of the diquark amplitude:

\begin{eqnarray}
\label{9}
A_1(s,s_{12})=
\frac{\alpha_1(s,s_{12}) B_1(s_{12})}{1-B_1(s_{12})} \, .
\end{eqnarray}

The equation for the reduced amplitude $\alpha_1(s,s_{12})$ can be written as

\begin{eqnarray}
\label{10}
\alpha_1(s,s_{12})&=&\lambda+\frac{1}{B_1(s_{12})}
\int\limits_{4m^2}^{\infty}
\, \frac{ds'_{12}}{\pi}\frac{\rho_1(s'_{12})}
{s'_{12}-s_{12}}G_1(s'_{12})\int\limits_{-1}^{+1} \, \frac{dz}{2}
\, \frac{2\alpha_1(s,s'_{13}) B_1(s'_{13})}{1-B_1(s'_{13})} \, .
\end{eqnarray}

The next step is to include into (\ref{10}) a cutoff at large $s'_{12}$. This
cutoff is needed to approximate the contribution of the interaction at
short distances. In this connection we shall rewrite Eq. (\ref{10}) as

\begin{eqnarray}
\label{11}
\alpha_1(s,s_{12})&=&\lambda+\frac{1}{B_1(s_{12})}
\int\limits_{4m^2}^{\infty}
\, \frac{ds'_{12}}{\pi}\Theta(\Lambda-s'_{12})\frac{\rho_1(s'_{12})}
{s'_{12}-s_{12}}G_1\int\limits_{-1}^{+1} \, \frac{dz}{2}
\, \frac{2\alpha_1(s,s'_{13}) B_1(s'_{13})}{1-B_1(s'_{13})} \, .
\end{eqnarray}

The construction of the approximate solution of Eq. (\ref{11}) is based on
extraction of the leading singularities are close to the region
$s_{ik}\approx 4m^2$. The structure of the singularities of amplitudes
with a different number of rescattering is the following \cite{26}.
The strongest singularities in $s_{ik}$ arise from pair rescatterings of
quarks: square-root singularity corresponding to a threshold and pole
singularities corresponding to bound states. We use the approximation in which
the singularity corresponding to a single interaction of all three particles,
the triangle singularity, is taken into account. Such a classification of
singularities makes it possible to search for an approximate solution
of Eq. (\ref{11}), taking into account a definite number of leading
singularities and neglecting the weaker ones.

For fixed values of $s$ and $s'_{12}$ the integration is carried out
over the region of the variable $s'_{13}$ corresponding to a physical
transition of the current into three quarks (the physical region of
Dalitz plot). It is convenient to take the central point of this region,
corresponding to $z=0$, to determinate the function $\alpha_1(s,s_{12})$
and also the Chew-Mandelstam function $B_1(s_{12})$ at the point
$s_{12}=s_0=\frac{s}{3}+m^2$. Then the equation for the $\Delta$ isobar
takes the form

\begin{equation}
\label{12}
\alpha_1(s,s_0)=\lambda+I_{1,1}(s,s_0)\cdot 2\, \alpha_1(s,s_0)
\, , \end{equation}

\begin{equation}
\label{13}
I_{1,1}(s,s_0)=\int\limits_{4m^2}^{\Lambda_1}
\, \frac{ds'_{12}}{\pi} \frac{\rho_1(s'_{12})}
{s'_{12}-s_{12}}G_1\int\limits_{-1}^{+1} \, \frac{dz}{2}
\, \frac{G_1}{1-B_1(s'_{13})}
\, . \end{equation}

We can obtain an approximate solution of Eq. (\ref{14})

\begin{equation}
\label{14}
\alpha_1(s,s_0)=\lambda [1-2\, I_{1,1}(s,s_0)]^{-1}
\, . \end{equation}

The function $I_{1,1}(s,s_0)$ takes into account correctly the singularities
corresponding to the fact that all propagators of triangle diagrams reduce to
zero. The right-hand side of (\ref{14}) may
have a pole in $s$, which corresponds to a bound state of the three quarks.
The choice of the cutoff $\Lambda$ makes it possible to fix the value of
the $\Delta$ isobar mass.

Baryons of $S$-wave multiplets have a completely symmetric spin-flavor
part of the wave function, and spin $\frac{3}{2}$ corresponds to the
decuplet which has a symmetric flavor part of the wave function. Octet
states have spin $\frac{1}{2}$ and a mixed symmetry of the flavor function.

In analogy with the case of the $\Delta$ isobar we can obtain the
rescattering amplitudes for all $S$-wave states with $J^P=\frac{3}{2}^+$,
which include quarks of various flavors. These amplitudes will satisfy
systems of integral equations. In considering the $J^P=\frac{1}{2}^+$
octet we must include the integration of the quarks in the $0^+$ and $1^+$
states (in the colour state $\bar 3_c$). Including all possible rescattering
of each pair of quarks and grouping the terms according to the final states
of the particles, we obtain the amplitudes $A_0$ and $A_1$, which satisfy
the corresponding systems of integral equations. If we choose the
approximation in which two-particle and triangle singularities are taken
into account, and if all functions which depend on the physical region of
the Dalitz plot, the problem of solving the system of integral equations
reduces to one of solving simple algebraic equations.

In our calculation the quark masses $m_u=m_d=m$ and $m_s$, are not uniquely
determined. In order to fix $m$ and $m_s$, anyhow, we make the simple
assumption that $m=\frac{1}{3} m_{\Delta} (1232)$,
$m_s=\frac{1}{3} m_{\Omega} (1672)$. The strange quark breaks the flavor
$SU(3)_f$ symmetry (\ref{6}).

In Ref. \cite{18} we consider two versions of calculations. In the first version
the $SU(3)_f$ symmetry is broken by the scale shift of the dimensional
parameters. A single cutoff parameter in pair energy is introduced for all
diquark states $\lambda_1=12.2$.

In the Table \ref{tab1} the calculated masses of the $S$-wave baryons are shown.
In the first version, we use only three parameters: the subenergy cutoff
$\lambda$ and the vertex function $g_0$, $g_1$, which corresponds to the
quark-quark interaction in $0^+$ and $1^+$ states. In this case, the mass
values of strange baryons with $J^P=\frac{1}{2}^+$ are less than the
experimental ones. This means that the contribution color-magnetic is too
large. In the second version, we introduce four parameters: cutoff
$\lambda_0$, $\lambda_1$ and the vertex function $g_0$, $g_1$. We decrease
the color-magnetic interaction in $0^+$ strange channels and calculated mass
values of two baryonic multiplets  $J^P=\frac{1}{2}^+$, $\frac{3}{2}^+$
are in good agreement with the experimental data \cite{27}.

The essential difference between $\Sigma$ and $\Lambda$ is the spin of the
lighter diquark. The model explains both the sign and magnitude of this mass
splitting.

The suggested method of the approximate solution of the relativistic
three-quark equations allows us to calculate the $S$-wave baryons mass
spectrum. The interaction constants, determined the baryons spectrum in
our model, are similar to ones in the bootstrap quark model of $S$-wave
mesons \cite{25}. The diquark interaction forces are defined by the gluon
exchange. The relative contribution of the instanton-induced interaction is
less than that with the gluon exchange. This is the consequence of
$1/N_c$-expansion.

The gluon exchange corresponds to the color-magnetic interaction, which
is responsible for the spin-spin splitting in the hadron models. The sign
of the color-magnetic term is such as to made any baryon of spin
$\frac{3}{2}$ heavier than its spin-$\frac{1}{2}$ counterpart (containing
the same flavors).

\begin{table}
\caption{$S$-wave baryon masses $M(J^p)$ $(GeV)$.}
\label{tab1}
\begin{tabular}{|ccp{4cm}ccc|}
\hline
 & $M(\frac{1}{2}^+)$ &  & & $M(\frac{3}{2}^+)$ &
\\[5pt]
\hline
$N$ &
\begin{tabular}{c}
$0.940$ \\
$0.940$
\end{tabular}
& $(0.940)$ & $\Delta$ &
\begin{tabular}{c}
$1.232$ \\
$1.232$
\end{tabular}
& $(1.232)$
\\[5pt]
$\Lambda$ &
\begin{tabular}{c}
$1.022$ \\
$1.098$
\end{tabular}
& $(1.116)$ & $\Sigma^*$ &
\begin{tabular}{c}
$1.377$ \\
$1.377$
\end{tabular}
& $(1.385)$
\\[5pt]
$\Sigma$ &
\begin{tabular}{c}
$1.050$ \\
$1.193$
\end{tabular}
& $(1.193)$ & $\Xi^*$ &
\begin{tabular}{c}
$1.524$ \\
$1.524$
\end{tabular}
& $(1.530)$
\\[5pt]
$\Xi$ &
\begin{tabular}{c}
$1.162$ \\
$1.325$
\end{tabular}
& $(1.315)$ & $\Omega$ &
\begin{tabular}{c}
$1.672$ \\
$1.672$
\end{tabular}
& $(1.672)$
\\[5pt]
\hline
\end{tabular}
\end{table}

\section{Nine-quark amplitudes of $ ^3 He$ $( ^3 H)$.}

We derive the relativistic nine-quark equations in the framework of the
dispersion relation technique. We use only planar diagrams; the other
diagrams due to the rules of $1/N_c$ expansion \cite{28, 29, 30} are neglected.
The current generates a nine-quark system. The correct equations for the
amplitude are obtained by taking into account all possible subamplitudes.
It corresponds to the division of complete system into subsystems with a
smaller number of particles. Then one should represent a nine-particle
amplitude as a sum of 36 subamplitudes:

\begin{eqnarray}
\label{15}
A=\sum\limits_{i<j \atop i, j=1}^9 A_{ij}\, . \end{eqnarray}

This defines the division of the diagrams into groups according to the
certain pair interaction of particles. The total amplitude can be
represented graphically as a sum of diagrams. We need to consider only
one group of diagrams and the amplitude corresponding to them, for example
$A_{12}$. We shall consider the derivation of the relativistic generalization
of the Faddeev-Yakubovsky approach.

In our case, the $ppn$ and $pnn$ states ($ ^3 He$ and $ ^3 H$) are considered.
The isospin of $ ^3 He$ is equal to $I=\frac{1}{2}$ and the spin-parity
$J^P=\frac{1}{2}^+$. We take into account the pairwise interaction of all
nine quarks in the nonaquark. The set of diagrams associated with the
amplitude $A_{12}$ can further be broken down into 15 groups corresponding
to subamplitudes. For instance, we consider
$A_1^{1^{uu}}(s,s_{12345678},s_{1234567},s_{123456},s_{12345},s_{1234},s_{123},s_{12})$.
Here $s_{ik}$ is the two-particle subenergy squared, $s_{ijk}$ corresponds
to the energy squared of particles $i$, $j$, $k$, $s_{ijk\ldots}$ is the energy
squared of particles $i$, $j$, $k$, \ldots and $s$ is the system
total energy squared.

In order to represent the subamplitudes $A_i$ in the form of a dispersion relation,
it is necessary to define the amplitude of $qq$ interactions. It is similar to
the three-quark case (Sec. II). We use the results of our relativistic quark
model \cite{23}

\begin{equation}
\label{16}
a_n(s_{ik})=\frac{G^2_n(s_{ik})}
{1-B_n(s_{ik})} \, ,\end{equation}

\begin{equation}
\label{17}
B_n(s_{ik})=\int\limits_{4m^2}^{\Lambda}
\, \frac{ds'_{ik}}{\pi}\frac{\rho_n(s'_{ik})G^2_n(s'_{ik})}
{s'_{ik}-s_{ik}} \, .\end{equation}

\noindent
Here $G_n(s_{ik})$ are the diquark vertex functions. The vertex
functions are determined by the contribution of the crossing channels.
The vertex functions satisfy the Fierz relations. These vertex
functions are generated from $g_V$. $B_n(s_{ik})$ and $\rho_n (s_{ik})$
are the Chew-Mandelstam functions with cutoff $\Lambda$ \cite{24}
and the phase spaces:

\begin{eqnarray}
\label{18}
\rho_n (s_{ik},J^P)=\left(\alpha(n,J^P) \frac{s_{ik}}{4m^2}
+\beta(n,J^P)\right)
\sqrt{\frac{s_{ik}-4m^2}{s_{ik}}}\, .
\end{eqnarray}

The coefficients $\alpha(n,J^P)$, $\beta(n,J^P)$ and
diquark vertex functions are given in Table \ref{tab2}.
Here $n=1$ coresponds to $qq$-pairs with $J^P=0^+$, $n=2$ corresponds
to the $qq$ pairs with $J^P=1^+$. In the case in question the interacting
quarks do not produce a bound states, thefore the integration in Eqs. (\ref{17})
is carried out from the threshold $4m^2$ to the cutoff $\Lambda$.

\begin{table}
\caption{Vertex functions and Chew-Mandelstam coefficients.}\label{tab2}
\begin{tabular}{|c|c|c|c|c|}
\hline
$n$ & $J^P$ & $G_n^2(s_{kl})$ & $\alpha_n$ & $\beta_n$ \\
\hline
& & & & \\
1 & $0^+$ & $\frac{4g}{3}-\frac{8gm^2}{(3s_{kl})}$
& $\frac{1}{2}$ & $0$ \\
& & & & \\
2 & $1^+$ & $\frac{2g}{3}$ & $\frac{1}{3}$
& $\frac{1}{6}$ \\
& & & & \\
\hline
\end{tabular}
\end{table}

Let us extract singularities in the coupled equations and obtain
the reduced amplitudes $\alpha_i$.

The reduced amplitudes $\alpha_1$ are determined by the channels
$1^{uu}$, $0^{ud}$, $1^{dd}$, the $\alpha_2$ are constructed as
$1^{uu}1^{uu}$, $1^{uu}0^{ud}$, $0^{ud}0^{ud}$, $1^{uu}1^{dd}$, $1^{dd}0^{ud}$,
$1^{dd}1^{dd}$ channels. The reduced amplitudes $\alpha_3$ are just the
$ppn$ states: $1^{uu}1^{uu}0^{ud}$, $1^{uu}1^{uu}1^{dd}$, $1^{uu}0^{ud}0^{ud}$,
$1^{uu}0^{ud}1^{dd}$, $0^{ud}0^{ud}0^{ud}$, $0^{ud}0^{ud}1^{dd}$.

We consider the 15 coupled equations. The contributions of two diquarks
and other 5 quarks mix with the $pp$ and $pn$ +3 quarks contributions.
The $\alpha_3$ do not include the $\alpha_3^{1^{uu}1^{dd}1^{dd}}$ state.

We used the classification of singularities, which was proposed in
paper \cite{26}. The construction of the approximate solution of
equation for the $\alpha_i$ is based on extraction of the leading singularities
of the amplitudes. The main singularities in $s_{ik}=4m^2$ are from pair
rescattering of the particles $i$ and $k$. First of all there are threshold
square-root singularities. Also possible are pole singularities which
correspond to the bound states. We take into account more weaker
singularities.

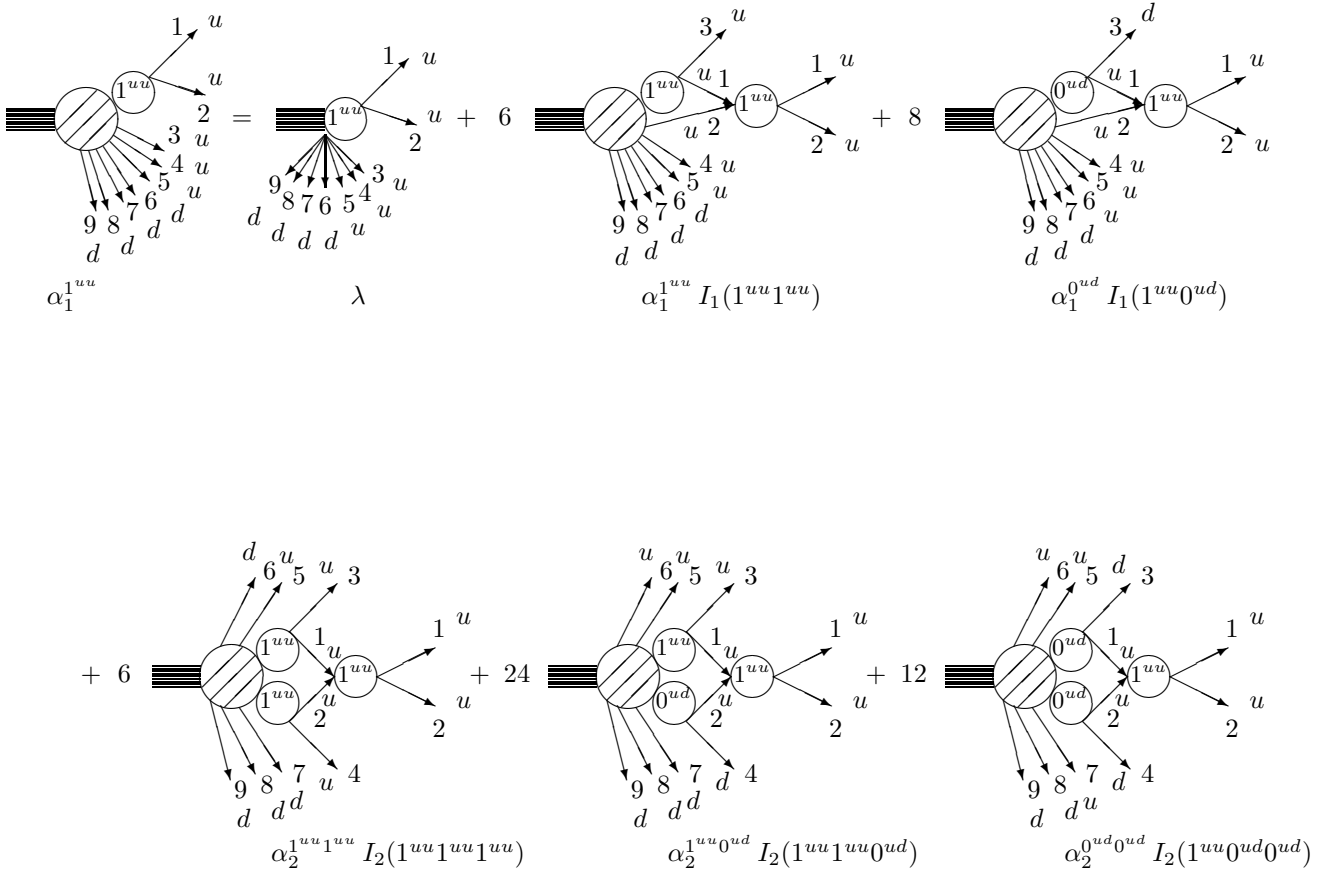
In the Fig. 2 the reduced amplitude $\alpha_1^{1^{uu}}$ is given.

The system of graphical equations in the Fig. 2 determines by the subamplitudes
using the self-consistent method. The coefficients are determined by the
permutation of quarks \cite{31, 32}. We should use the coefficients
multiplying of the diagrams in the graphical equation Fig. 2.

In the Fig. 2 the first coefficient is equal to 6, that the number $6=2$
(permutation particles 1 and 2) $\times 3$ (we can use third, 4-th, 5-th
$u$-quarks); the second coefficient equal to 8, that the number $8=2$
(permutation particles 1 and 2) $\times 4$ (we can use third, 4-th, 5-th,
6-th $d$-quarks); the third coefficient equal to 6: $6=3$ (we can replace third
$u$-quark with any of two remained $u$-quarks) $\times 2$ (then replace 4-th
$u$-quark with remained $u$-quark); the 4-th coefficient equal to 24:
$24=2$ (we can replace third $u$-quark with 4-th $d$-quark) $\times 3$
(we can replace third $u$-quark with any of two remained $u$-quarks)
$\times 4$ (we can replace 4-th $d$-quark with any of three remained $d$-quarks);
the 5-th coefficient equal to 12: $12=4$ (we can replace third $d$-quark with
any of three remained $d$-quarks) $\times 3$ (we can replace 4-th $d$-quark
with any of two remained $d$-quarks).

The similar approach allows us to take into account the coefficients
in all equations.

With this classification, one defines the reduced amplitudes $\alpha_1$, $\alpha_2$,
$\alpha_3$ as well as the $B$-functions in the middle point of physical region
of Dalitz-plot at the point $s_0$:

\begin{eqnarray}
\label{19}
s_0=\frac{s+63}{36} \, , \quad s_{ijk}=3s_0-3 \, , \quad
s_{ijkl}=6s_0-8 \, ,
\end{eqnarray}

\begin{eqnarray}
\label{20}
s=\sum\limits_{i,j=1 \atop i<j}^{9} s_{ij}-63m^2 \, , \quad
s_{ij}=s_0 m^2 \, , \quad s=\left (\frac{M}{m}\right )^2 \, .
\end{eqnarray}

Such choice of point $s_0$ allows us to replace integral equations
of Fig. 2 by the algebraic equations, for instance:

\begin{eqnarray}
\label{21}
\alpha_1^{1^{uu}}&=&\lambda+6\, \alpha_1^{1^{uu}} I_1(1^{uu}1^{uu})
+8\, \alpha_1^{0^{ud}} I_1(1^{uu}0^{ud})
+6\, \alpha_2^{1^{uu}1^{uu}} I_2(1^{uu}1^{uu}1^{uu})
+24\, \alpha_2^{1^{uu}0^{ud}} I_2(1^{uu}1^{uu}0^{ud})
\nonumber\\
&&\nonumber\\
&+&12\, \alpha_2^{0^{ud}0^{ud}} I_2(1^{uu}0^{ud}0^{ud}) \, .
\end{eqnarray}

\begin{eqnarray}
\label{22}
I_1(ij)&=&\frac{B_j(s_0^{13})}{B_i(s_0^{12})}
\int\limits_{4m^2}^{\Lambda}
\frac{ds'_{12}}{\pi}\frac{G_i^2(s_0^{12})\rho_i(s'_{12})}
{s'_{12}-s_0^{12}} \int\limits_{-1}^{+1} \frac{dz_1(1)}{2}
\frac{1}{1-B_j (s'_{13})}\, , \\
&&\nonumber\\
\label{23}
I_2(ijk)&=&\frac{B_j(s_0^{13}) B_k(s_0^{24})}{B_i(s_0^{12})}
\int\limits_{4m^2}^{\Lambda}
\frac{ds'_{12}}{\pi}\frac{G_i^2(s_0^{12})\rho_i(s'_{12})}
{s'_{12}-s_0^{12}}
\frac{1}{2\pi}\int\limits_{-1}^{+1}\frac{dz_1(2)}{2}
\int\limits_{-1}^{+1} \frac{dz_2(2)}{2}\nonumber\\
&&\nonumber\\
&\times&
\int\limits_{z_3(2)^-}^{z_3(2)^+} dz_3(2)
\frac{1}{\sqrt{1-z_1^2(2)-z_2^2(2)-z_3^2(2)+2z_1(2) z_2(2) z_3(2)}}
\nonumber\\
&&\nonumber\\
&\times& \frac{1}{1-B_j (s'_{13})} \frac{1}{1-B_k (s'_{24})}
 \, ,
\end{eqnarray}

\noindent
where $i$, $j$, $k$ correspond to the diquarks with the
spin-parity $J^P=0^+$, $1^+$.

The other choices of point $s_0$ do not change essentially the contributions
of $\alpha_l$, $l=1-3$, therefore we omit the indices $s_0^{ik}$.

We calculate the realistic case of $ ^3 He$, containing the $u$ and $d$
quarks. The solutions of the system of 15 equations are considered as
$\alpha_i(s)=F_i(s,\lambda_i)/D(s)$, where zeros of $D(s)$ determinant
define the mass of bound states ($ ^3 He$ and $ ^3 H$).

The mass of $ ^3 He$ is equal to $M=2809\, MeV$. We propose
the masses of $p$ and $n$ states are equal to $M=940\, MeV$
(without electromagnetic interactions), then the $ ^3 H$ mass is
equal to the mass of $ ^3 He$.

The reduced amplitudes of $ ^3 He$ are given in the Appendix A.

\section{Calculation results.}

The poles of the reduced amplitudes $\alpha_l$ ($l=1-3$) correspond to the
bound state and determines the mass of the nine-quark state with $J^P=\frac{1}{2}^+$
and isospin $I=\frac{1}{2}$. The quark $u$, $d$ masses are equal to $m_{u,d}=410\, MeV$,
coincide with the ordinary baryon ones in our model (Sec. II). The model in question
has only two parameters: the cutoff parameter $\Lambda=11$ (similar to the three quark
model (Sec. II)) and the gluon coupling constant $g_V=0.1536$. This parameter
is determined by the $ ^3 He$ mass $M=2809\, MeV$. The $ ^3 H$
mass is equal to one. In the nuclei region the gluon interaction is smaller
in 2 -- 3 time as compared of the gluon interaction of low-lying hadrons (0.35 -- 0.45).

The estimation of theoretical error on the $ ^3 He$ is equal to
$1\, MeV$. This results was obtained by the choice of model parameters.
We predict the mix $pn$ + 3 quarks, $pp$ + 3 quarks systems and two
diquark + 5 quarks.

The functions $I_1$, $I_2$, $I_3$, $I_4$, $I_5$, $I_6$, $I_7$, $I_8$,
$I_9$, $I_{10}$ are taken in paper \cite{19}. The other functions are
determined by the following formulae (see also Fig. 3):

\begin{eqnarray}
\label{24}
I_{11}(12,34,15,36,47)=I_1(12,15)\times I_2(34,36,47) \, ,
\end{eqnarray}

\begin{eqnarray}
\label{25}
I_{12}(12,34,56,17)=I_1(12,17) \, ,
\end{eqnarray}

\begin{eqnarray}
\label{26}
I_{13}(12,34,56,17,28)=I_2(12,17,28) \, ,
\end{eqnarray}

\begin{eqnarray}
\label{27}
I_{14}(12,34,56,17,38)=I_1(12,17)\times I_1(34,38) \, ,
\end{eqnarray}

\begin{eqnarray}
\label{28}
I_{15}(12,34,56,23,47)=I_7(12,34,23,47) \, ,
\end{eqnarray}

\begin{eqnarray}
\label{29}
I_{16}(12,34,56,17,23,48)=I_8(12,34,17,23,48) \, ,
\end{eqnarray}

\begin{eqnarray}
\label{30}
I_{17}(12,34,56,17,45)=I_1(12,17)\times I_3(34,56,45) \, ,
\end{eqnarray}

\begin{eqnarray}
\label{31}
I_{18}(12,34,56,17,38,49)=I_1(12,17)\times I_2(34,38,49) \, ,
\end{eqnarray}

\begin{eqnarray}
\label{32}
I_{19}(12,34,56,17,38,59)=I_1(12,17)\times I_1(34,38)\times I_1(56,59) \, ,
\end{eqnarray}

\begin{eqnarray}
\label{33}
I_{20}(12,34,56,17,45,68)=I_1(12,17)\times I_7(34,56,45,68) \, ,
\end{eqnarray}

\begin{eqnarray}
\label{34}
I_{21}(12,34,56,17,28,45)=I_2(12,17,28)\times I_3(34,56,45) \, .
\end{eqnarray}

For the new function $I_{22}$ we used the following equation:

\begin{eqnarray}
\label{35}
I_{22}(ijklmn)&=&\frac{B_l(s_0^{23})B_m(s_0^{45})B_n(s_0^{67})}
{B_i(s_0^{12}) B_j(s_0^{34}) B_k(s_0^{56})}
\int\limits_{4m^2}^{\Lambda}
\frac{ds'_{12}}{\pi}\frac{G_i^2(s_0^{12})\rho_i(s'_{12})}{s'_{12}-s_0^{12}}
\int\limits_{4m^2}^{\Lambda}
\frac{ds'_{34}}{\pi}\frac{G_j^2(s_0^{34})\rho_j(s'_{34})}{s'_{34}-s_0^{34}}
\nonumber \\
&&\nonumber\\
&\times &
\int\limits_{4m^2}^{\Lambda}
\frac{ds'_{56}}{\pi}\frac{G_k^2(s_0^{56})\rho_k(s'_{56})}{s'_{56}-s_0^{56}}
\frac{1}{(2\pi)^2}
\int\limits_{-1}^{+1} \frac{dz_1}{2}
\int\limits_{-1}^{+1} \frac{dz_2}{2}
\int\limits_{-1}^{+1} \frac{dz_3}{2}
\int\limits_{-1}^{+1} \frac{dz_4}{2}
\int\limits_{-1}^{+1} \frac{dz_6}{2}
\int\limits_{-1}^{+1} \frac{dz_7}{2}
\nonumber \\
&&\nonumber\\
&\times &
\int\limits_{z_5^-}^{z_5^+} dz_5
\int\limits_{z_8^-}^{z_8^+} dz_8
\frac{1}{\sqrt{1-z_3^2-z_4^2-z_5^2+2z_3z_4z_5}}
\frac{1}{\sqrt{1-z_2^2-z_7^2-z_8^2+2z_2z_7z_8}}
\nonumber \\
&&\nonumber\\
&\times &
\frac{1}{1-B_l (s'_{23})}\frac{1}{1-B_m (s'_{45})}
\frac{1}{1-B_n (s'_{67})}\, .
\end{eqnarray}

The contribution of this new functions is about $10^{-8}$ and
the contribution of $I_{10}$, $I_{17}$, $I_{19}$, $I_{20}$, $I_{21}$
(Eqs. (\ref{30}), (\ref{32}) -- (\ref{34})) are similar to $I_{22}$.
We do not take into account these functions in Appendix A.
The $ijklmn$ determine the diquarks contributions.

\section{Conclusions.}

The structure and interactions of the light nuclei have been the focus
of experimental and theoretical exploration since the infancy of nuclear
physics.

The known way with which to calculate the low-energy properties of
hadronic and nuclear systems rigorously is Lattice QCD (LQCD). In LQCD
calculations, the quark and gluon fields are defined on a discretized
space-time of finite volume of the lattice, such deviations can
be systematically removed by reducing the lattice spacing, increasing
the lattice volume and extrapolating to the continuum and infinite volume
limits using the known dependences determined with effective field theory
(EFT). Calculation of important quantifies in nuclear physics using LQCD
is only now becoming practical, with first calculations of simple
multi-baryon interactions being recently performed, although not at the
physical values of the light-quark masses. Early exploratory quenched
calculations of the $NN$ scattering lengths \cite{35, 36} performed more than
a decade ago have been used by $n_f=2+1$ calculations within the last few
years \cite{37, 38} (and added to by futher quenched calculations \cite{39, 40}).

Further, the first quenched calculations of the deuteron \cite{35}, $ ^3 He$
and $ ^4 He$ \cite{36} have been performed with $n_f=2+1$ calculations of
$ ^3 He$ and multi-baryon systems containing strange quarks \cite{37}.
In addition efforts to explore nuclei and nuclear matter using the strong
coupling limit of QCD have led to same interesting observations \cite{38}.
Future calculation at smaller spacing and at lighter quark masses will
facilitate such extrapolations and lead to first predictions for the spectrum
of light nuclei, with completely quantified uncertainties, that can be
compared with experiment.

We have considered the dibaryons with the $u$, $d$, $s$-quarks.
The $H$-particle, $N\Omega$-state and di-$\Omega$ may be strong interaction
stable. Up to now, these three interesting candidates of dibaryons are still
not found or confirmed by experiments. It seems that one should go beyond
these candidates and should search the possible candidates in a wider
region, especially the systems with multistrangeness in terms of a more
reliable model.

In our paper the dynamics of quark interactions is defined by the
Chew-Mandelstam functions (Table \ref{tab2}). We include only two
parameters: the cutoff $\Lambda$, gluon coupling constant $g_V$.

The comparison of baryons with the $ ^3 He$ and $ ^3 H$ state gluon
coupling constants gives rise to 2 -- 3 time smaller the baryon one.

The quark-gluon interaction
allows us to calculate the mass spectrum at light hypernuclei.

\begin{acknowledgments}
The authors would like to thank to T. Barnes, L.V. Krasnov for useful discussion.
\end{acknowledgments}

\newpage



\newpage

\appendix

\section{}

The coupled equations for the reduced amplitudes of $ ^3 He$:

%

\end{document}